\documentclass[11pt,twoside]{article}
\usepackage{asp2010}
\usepackage{rotating}

\resetcounters

\begin{document}

\title{Analytical Representation for Equations of State of Dense Matter}

\author{C.\ G\"{u}ng\"{o}r, K. Y. Ek\c{s}i}
\affil{Istanbul Technical University}

\begin{abstract}
We present an analytical unified representation for 22 equations of state (EoS)
of dense matter in neutron stars.
Such analytical representations can be useful for modeling neutron star structure 
in modified theories of gravity with high order derivatives.
\end{abstract}

\section{Introduction}
The different possible equations of state (EoS) of dense matter inside neutron stars (NSs) are often
given as tabulated data which are fed into the solution of the hydrostatic equilibrium equations 
via an interpolation technique. While this method works well in the case of general relativity (GR)
for which the hydrostatic equilibrium is described by the TOV equations \citep{tol39,opp39},
it leads to problems in modified theories of gravity, e.g.\ $f(R)$ theories \citep{def10} in which 
the equations describing the hydrostatic equilibrium 
contain second derivatives of $P(\rho)$
(see e.g.\ \citet{ara11}) where $P$ is the pressure and $\rho$ is the density. In such cases the order of the 
polynomial used in interpolation technique may effect the results if it is not sufficiently high
and the linear interpolation technique will certainly fail. 
It is then necessary to employ analytical representations of the EoS that are
sufficiently many times differentiable. In this work, we provide an accurate unified analytical 
representation for 22 EoS.

\section{Analytical Representation of EoS}

A unified analytical representation for two EoS, FPS 
and SLY, were provided 
by \citet{HP04} (hereafter HP04) where the authors provide a relation between  $\zeta =\log(P /\mathrm{dyn\, cm^{-2}})$ 
and  $\xi =\log(\rho /\mathrm{g\, cm^{-3}})$ 
with 18 free parameters.
The unified representation we provide here is an extension of HP04 with 23 parameters 12 of which 
are fixed to represent low density regimes of BPS \citep{ref_BPS} and NV \citep{ref_NV}. 
The rest 11 free parameters are used to fit the different EoS at high density regimes and to match them.
The function we use to represent EoS for NS is
\begin{equation}
\zeta =\zeta _{\mathrm{low}}f_0( a_1 ( \xi -c_{11})) 
     +f_0( a_2( c_{12}-\xi ) ) \zeta _{\mathrm{high}}
 \label{eq.eos}    
\end{equation}
where $f_{0}(x)=1/(1+\exp x)$ is the matching function (also used in HP04).
\pagebreak
\newpage
\noindent Here
\begin{equation}
\zeta _{\mathrm{low}} =\left[ c_1+c_2( \xi -c_3)^{c_4}\right] f_0( c_5( \xi -c_6) ) 
                      +(c_7+c_8\xi )f_0( c_9( c_{10}-\xi ))
\end{equation}  
and   
\begin{equation}                 
\zeta _{\mathrm{high}} = (a_{3}+a_{4}\xi )f_{0}\left( a_{5}\left( a_{6}-\xi
\right) \right)  
+(a_{7}+a_{8}\xi +a_{9}\xi ^{2})f_{0}\left( a_{10}\left( a_{11}-\xi
\right) \right)                       
\end{equation}
describe the low and high density regimes, respectively. 
We provide the values of the fit parameters $c_i$ and $a_i$ for $\xi>5$ in  Table~\ref{tab:fit.P.c} and Table~\ref{tab:fit.P.a}.


\begin{figure}[t]
\begin{minipage}[b]{0.60\textwidth}
\centering
\includegraphics[width=0.9\textwidth]{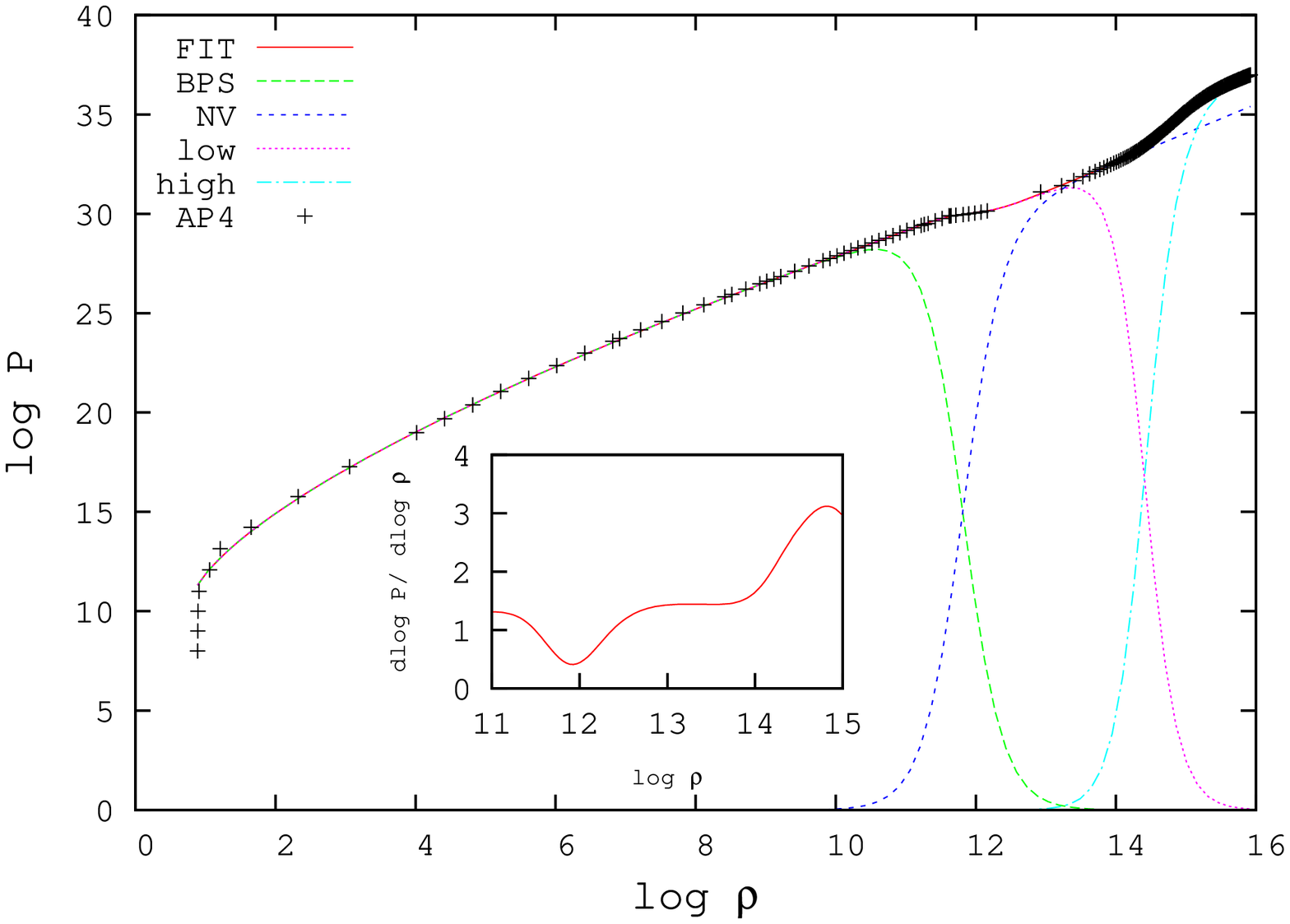}
\caption{The EoS AP4 (+) and its analytical representation (red line). 
The purple dotted line and turquoise dash-dotted line describe low and high density regimes, respectively.
The small figure shows the adiabatic index $\Gamma=d\zeta/d\xi$ versus $\xi$.}
\label{fig_AP4_fit}
\end{minipage}
\begin{minipage}[b]{0.40\textwidth}
\centering
\includegraphics[width=0.9\textwidth]{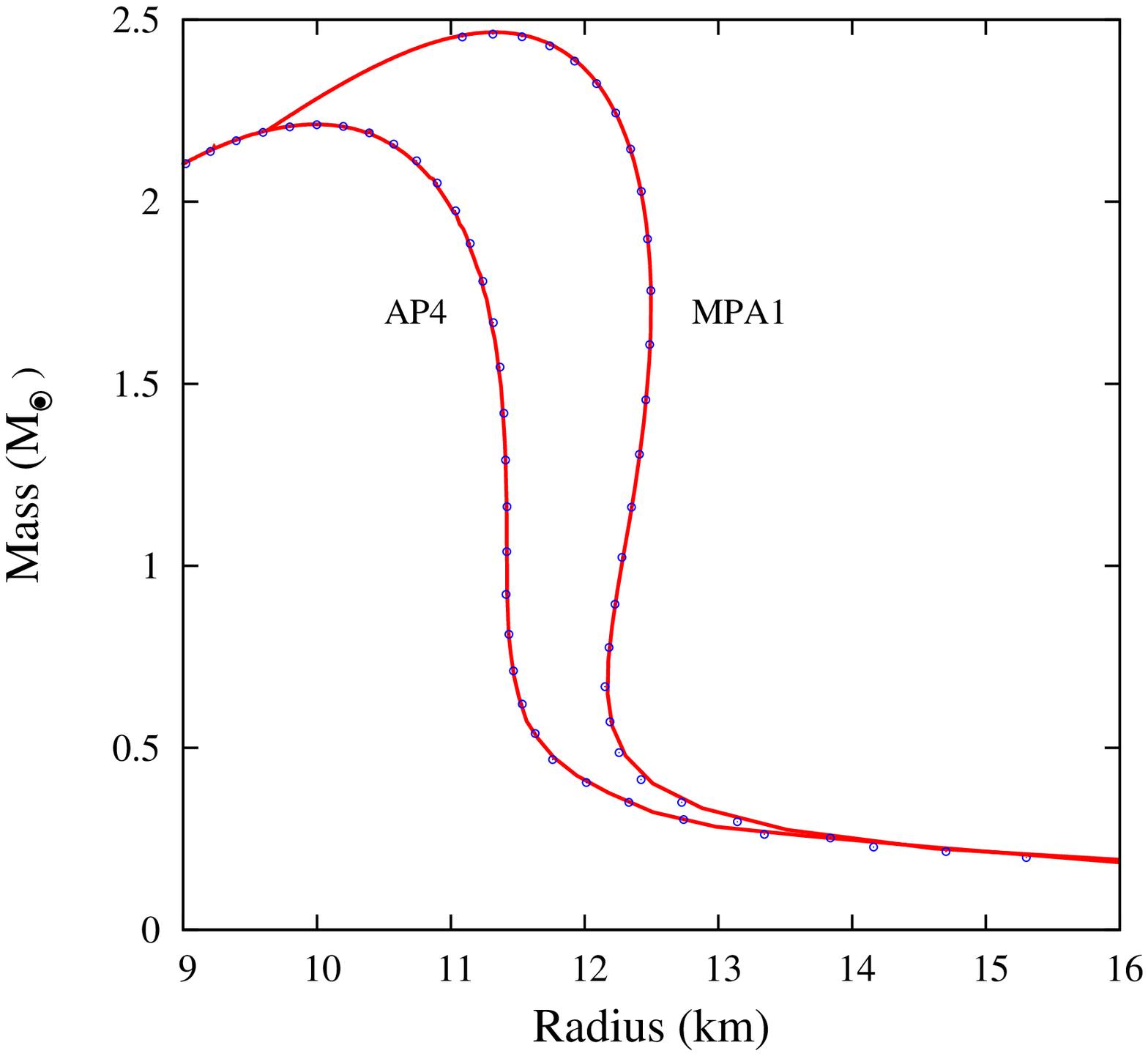}
\caption{M-R relation for EoS, AP4 and MPA1.  The red line is obtained by the analytical representation 
of the EoSs as presented in this work while the blue circles are obtained by the tabulated EoS.}
\label{fig_MR}
\end{minipage}
\end{figure}

In Figure~\ref{fig_AP4_fit}, we show our results for fitting the EoS data of AP4 with Equation~(\ref{eq.eos}).
In Figure~\ref{fig_MR}, we compare the M-R relations obtained by the analytical representation 
with that obtained by feeding the EoS data via interpolation technique.
The maximum relative error is $\sim 0.05\%$ near the maximum mass.

\section{Discussion and Conclusion}

The unified analytical expression presented in this work is an accurate representation 
of many EoS suggested for NSs. 
The analytical representation is preferable for solving the structure of NSs in modified theories of gravity where hydrostatic equilibrium 
equations are of 4th order. In such cases the usual interpolation technique will fail because the high order 
derivatives may not be continuous if the order 
of the polynomial used for interpolation is not sufficiently high.

\acknowledgements
We thank J.~M.\ Lattimer for providing the EoS data. We acknowledge support from the 
Scientific and Technological Council of TURKEY (TUBITAK) through grant number 108T686.

\pagebreak

\begin{table}
\centering
\caption{Parameters of the fit for low regimes}
\label{tab:fit.P.c}%
\begin{tabular}{r|l}
\hline\hline
\rule[-1.4ex]{0pt}{4.3ex} c$_1$ & 10.6557 \\ 
c$_2$ & 3.7863 \\ 
c$_3$ & 0.8124 \\ 
c$_4$ & 0.6823 \\ 
c$_5$ & 3.5279 \\ 
c$_6$ & 11.8100 \\ 
c$_7$ & 12.0584 \\ 
c$_8$ & 1.4663 \\ 
c$_9$ & 3.4952 \\ 
c$_{10}$ & 11.8007 \\ 
c$_{11}$ & 14.4114 \\ 
c$_{12}$ & 14.4081 \\ \hline\hline
\end{tabular}%
\end{table}

\begin{tiny}
\begin{sidewaystable} [t] \footnotesize
\caption{Parameters of the fit}
\centering 
\label{tab:fit.P.a}%
\begin{tabular}{r|l|l|l|l|l|l|l|l|l|l|l}
\hline\hline
\rule[-1.4ex]{0pt}{4.3ex} EoS & AP1 & AP2 & AP3 & AP4 & engvik & gm1nph & gm2nph & gm3nph & mpa1 & ms00 & ms2 
\\ \hline
\rule{0pt}{2.7ex} a$_1$ & 4.3290 & 4.3290 & 6.3293 & 4.3290 & 5.0487 & 
11.4832 & 10.9801 & 11.5163 & 5.2934 & 16.6491 & 14.0084 \\ 
a$_2$ & 4.3622 & 4.3622 & 6.3467 & 4.3622 & 5.0838 & 11.4006 & 10.9233 & 
11.4210 & 5.3319 & 16.2520 & 13.8422 \\ 
a$_3$ & 138.1760 & 184.1600 & 174.2030 & 9.1131 & -100.7550 & -100.7660 & 
-100.9130 & -100.8670 & 87.7901 & 8.0809 & 16.5970 \\ 
a$_4$ & -10.1093 & -13.4285 & -15.0960 & -0.4751 & 9.2091 & 11.8432 & 15.6165
& 16.0001 & -5.8466 & -0.5036 & -1.0943 \\ 
a$_5$ & 6.0097 & 5.8719 & 5.2453 & 3.4614 & 1.5946 & 1.2081 & 1.0707 & 0.9211
& 2.7232 & 6.8715 & 5.6701 \\ 
a$_6$ & 14.0120 & 13.8551 & 13.8962 & 14.8800 & 13.9137 & 12.7088 & 11.9067
& 11.7186 & 15.0804 & 14.7513 & 14.8169 \\ 
a$_7$ & -411.1380 & -372.6920 & -377.4200 & 21.3141 & 462.5180 & 462.0230 & 
461.2360 & 461.2870 & 428.4130 & -39.9099 & -56.3794 \\ 
a$_8$ & 48.0721 & 40.4186 & 44.3688 & 0.1023 & -48.4401 & -51.9392 & -55.7395
& -55.2969 & -57.6403 & 8.0314 & 9.6159 \\ 
a$_9$ & -1.1630 & -0.8061 & -0.8911 & 0.0495 & 1.1892 & 1.2438 & 1.2502 & 
1.2024 & 2.0957 & -0.2022 & -0.2332 \\ 
a$_{10} $ & 4.7514 & 4.7456 & 4.5062 & 4.9401 & 5.0896 & 5.0896 & 5.0897 & 
5.0897 & 5.0588 & 5.0102 & -3.8369 \\ 
a$_{11} $ & 10.7234 & 10.7503 & 13.6882 & 10.2957 & 10.2277 & 10.2277 & 
10.2276 & 10.2276 & 10.2727 & 10.3408 & 23.1860 \\ \hline
\rule{0pt}{2.7ex} & ms1506 & pal2 & pclnphq & wff1 & wff2 & wff3 & wff4 & 
schaf1 & schaf2 & prakdat & ps \\ \hline
\rule{0pt}{2.7ex} a$_1$ & 15.4744 & 11.3079 & 11.3180 & 2.8144 & 3.9694 & 
5.2934 & 5.0395 & 11.9204 & 13.3358 & 12.3641 & 12.3769 \\ 
a$_2$ & 15.1705 & 11.1963 & 11.2234 & 2.8539 & 4.0017 & 5.2912 & 5.0512 & 
11.7992 & 13.1285 & 12.2125 & 12.0277 \\ 
a$_3$ & 2.8727 & -0.3393 & -9.2769 & -6.6340 & -6.1548 & -5.8771 & -5.9160 & 
-11.7736 & -6.0249 & -9.1898 & -4.5181 \\ 
a$_4$ & -0.2014 & 0.0149 & 0.6194 & 9.3448 & 20.5349 & 29.4964 & 29.5087 & 
3.1670 & 2.7698 & 1.0249 & 2.6750 \\ 
a$_5$ & 4.2468 & 5.0914 & 7.9411 & 0.8728 & 0.5724 & 0.2267 & 0.1944 & 
-12.0189 & -13.5535 & 2.5784 & -11.4273 \\ 
a$_6$ & 15.1919 & 14.5066 & 15.2677 & 15.2163 & 15.3019 & 17.8721 & 21.5182
& 14.6274 & 14.7235 & 13.6433 & 14.5886 \\ 
a$_7$ & -83.9695 & -68.4757 & -194.2320 & -379.9680 & -377.0480 & -380.2820
& -381.1690 & 15.3263 & -398.4300 & 44.5552 & -84.9113 \\ 
a$_8$ & 13.8824 & 11.8013 & 28.9879 & 77.0628 & 85.8006 & 65.3388 & 61.2652
& 0.5165 & 55.1217 & -2.3817 & 13.9429 \\ 
a$_ 9$ & -0.3959 & -0.3266 & -0.9137 & -3.5619 & -4.5011 & -3.1750 & -2.6609
& 0.0521 & -1.7487 & 0.0902 & -0.3964 \\ 
a$_{10}$ & -4.3087 & -4.3089 & -4.3181 & -4.3181 & -4.3181 & -4.3181 & 
-4.3181 & 12.1722 & 13.6495 & 4.8729 & 11.5048 \\ 
a$_{11}$ & 23.3603 & 23.3604 & 23.3626 & 23.3626 & 23.3626 & 23.3626 & 
23.3626 & 14.6270 & 14.7214 & 10.3969 & 14.5874 \\ \hline\hline
\end{tabular}%
\end{sidewaystable}
\end{tiny}

\end{document}